\documentclass[journal=jacsat,manuscript=article,caption,float,geometry,natbib,setspace,xkeyval,maxauthors=0]{achemso}

\usepackage{amsmath}             
\usepackage{amssymb}             
\usepackage{wasysym}             
\usepackage{adjustbox}
\usepackage{natbib}
\usepackage{multirow}
\usepackage{multicol}
\usepackage{color}               
\usepackage{setspace}            
\usepackage{graphicx}            
\usepackage{wrapfig}             
\usepackage[dvipsnames]{xcolor}  
\usepackage{orcidlink}
\usepackage{subcaption}
\usepackage{hyperref} 
\usepackage[font=footnotesize,labelfont=bf,labelsep=period,width=0.75\textwidth]{caption}

\hypersetup{
    pdftitle={Multilayer Ferromagnetic Spintronic Devices for Neuromorphic Computing Applications},
   colorlinks=true,
    breaklinks=true,
    bookmarksopen=true,
    linkcolor=red,  
    citecolor=green, 
    urlcolor=blue   
}

\author{Aijaz H. Lone\orcidlink{https://orcid.org/0000-0002-1687-2917}}%
        \email{aijaz.lone@kaust.edu.sa}
        \affiliation{Division of Computer, Electrical and Mathematical Sciences and Engineering (CEMSE), King Abdullah University of Science and Technology (KAUST), Saudi Arabia }
\author{Xuecui Zou}
        \affiliation{Division of Computer, Electrical and Mathematical Sciences and Engineering (CEMSE), King Abdullah University of Science and Technology (KAUST), Saudi Arabia }
        \author{Kishan K. Mishra\orcidlink{https://orcid.org/0000-0002-5369-0880}}
         \affiliation{Department of Physics \& Astrophysics, University of Delhi, India}%

        \author{Venkatesh Singaravelu}
        \affiliation{Nanofabrication Core Lab, King Abdullah University of Science and Technology (KAUST), Saudi Arabia }

\author{Hossein Fariborzi}%
        \email{hossein.fariborzi@kaust.edu.sa}
        \affiliation{Division of Computer, Electrical and Mathematical Sciences and Engineering (CEMSE), King Abdullah University of Science and Technology (KAUST), Saudi Arabia }
        
\author{Gianluca Setti}%
        \email{gianluca.setti@kaust.edu.sa}
        \affiliation{Division of Computer, Electrical and Mathematical Sciences and Engineering (CEMSE), King Abdullah University of Science and Technology (KAUST), Saudi Arabia }
\title{Multilayer Ferromagnetic Spintronic Devices for Neuromorphic Computing Applications}

\begin{document}
\begin{abstract}
\setstretch{1} 
Spintronics has gone through substantial progress due to its applications in energy-efficient memory, logic and unconventional computing paradigms. Multilayer
ferromagnetic thin films are extensively studied for understanding the domain wall and skyrmion dynamics. However, most of these studies are confined to the materials and domain wall/skyrmion physics. In this paper, we present the experimental and micromagnetic realization of a multilayer ferromagnetic spintronic device for neuromorphic
computing applications. The device exhibits multilevel resistance states and the number of resistance states increases with lowering temperature. This is supported by the multilevel magnetization behavior observed in the micromagnetic simulations. Furthermore, the evolution of resistance states with spin-orbit torque is also explored in experiments and simulations. Using the multi-level resistance
states of the device, we propose its applications as a synaptic device in hardware
neural networks and study the linearity performance of the synaptic devices. The neural network based on these devices is trained and tested on the MNIST dataset using a supervised
learning algorithm. The devices at the chip level achieve 90\% accuracy. Thus, proving its applications in neuromorphic computing. Furthermore, we lastly discuss
the possible application of the device in cryogenic memory electronics for quantum
computers.
\end{abstract}
\begin{quotation}
\textit{keywords:\ {{Spintronics,  Domain wall devices, mutli- state memory, Micromagnetics, Synaptic devices and Neuromorphic computing,}}}
\end{quotation}

\section{Introduction}
Spintronic devices have gained a lot of interest for their application in high-integration data storage and energy-efficient computing applications \cite{1,2,3,4,5}. In particular, owing to the different operating behaviours such as binary-deterministic \cite{6,7,8}, stochastic \cite{9,10,11,12} and analog resistance \cite{13,14}, these devices are proving to be quite promising for neuromorphic computing applications. Magnetic domain wall and skyrmion devices as alternate analog memory \cite{15,16,17} and logic technology \cite{18,19} have been explored for their applications in neuromorphic computing \cite{20,21,22}. To stabilize the room temperature skyrmions and domains spintronic perpendicular magnetic anisotropy PMA-based magnetic multilayer systems have been extensively studied from material and physical perspective \cite{23,24,25,26}. These systems are used to study the domain wall and skyrmions dynamics in different magnetic materials and their interface with the other heavy metals \cite{27,28,29}. But to the best of our knowledge, not much effort has been made in employing these systems for neuromorphic computing. Considering the interesting physical phenomenon in these systems thus, the associated emerging device characteristics.   
In this paper, we present the experimental and micromagnetic realization of a spintronic device exhibiting discrete resistance states. The discreteness of the device behavior increases as we lower the temperature and for 100K temperature, 15 resistance states are observed. We attribute this discrete resistance behaviour to the magnetic domain wall pinning/depinning and gradual switching of different magnetic layers at low temperatures. The discrete resistance behavior is also observed in the micromagnetic simulations of similar crossbars of different widths. Furthermore, the evolution of resistance states with current pulses providing spin transfer and spin-orbit torque is also explored in experiments and simulations. Using the multi-level resistance state of the device, we propose its applications as a synaptic device in hardware neural networks and study the linearity performance of the synaptic devices. We map these resistance states to the weights of a neural network architecture. The network based on these devices is trained and tested on the MNIST dataset using a supervised learning algorithm. The system shows the accuracy performance up to 90\% which is comparable to the majority of the beyond CMOS synaptic devices. The discrete resistance states across the range of temperatures open a possible application of these devices in cryogenic electronics for quantum computers. 

\begin{figure}[!ht]
    \centering
    \includegraphics[width =\linewidth]{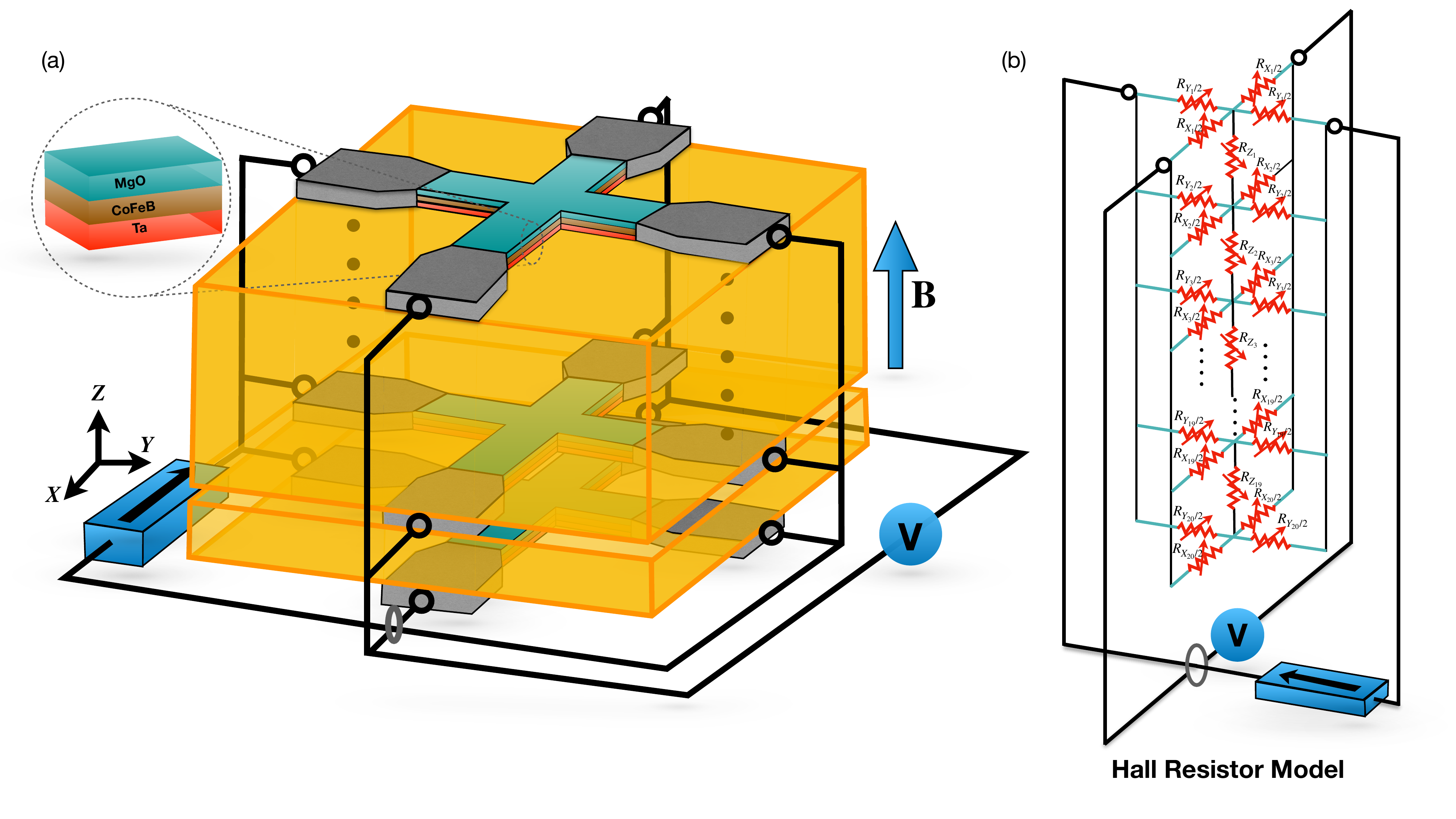}
    \caption{(a) Domain wall crossbar devices. (b) Equivalent resistor model.}
    \label{fig:Fig1}
\end{figure}
\section{Results and discussion}
\begin{figure}[htbp]
    \centering
    \includegraphics[width = \linewidth]{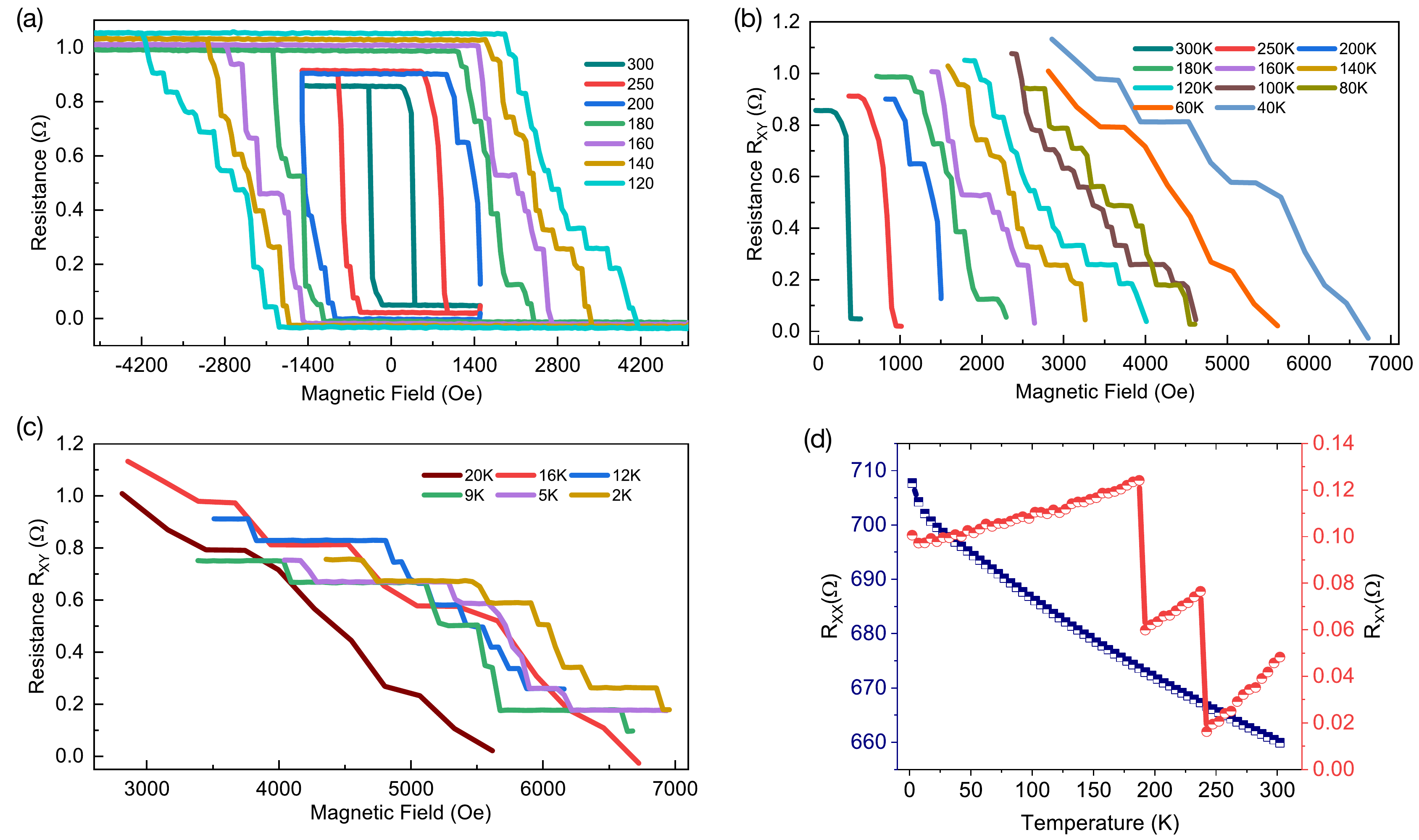}
    \caption{(a-c) Discrete Anomalous Hall resistance at different temperatures (2K to 300K). (d) Longitudinal and anomalous Hall resistance at different temperatures.}
    \label{fig:Fig2}
\end{figure}

\begin{figure}[t]
    \centering
    \includegraphics[width = \linewidth]{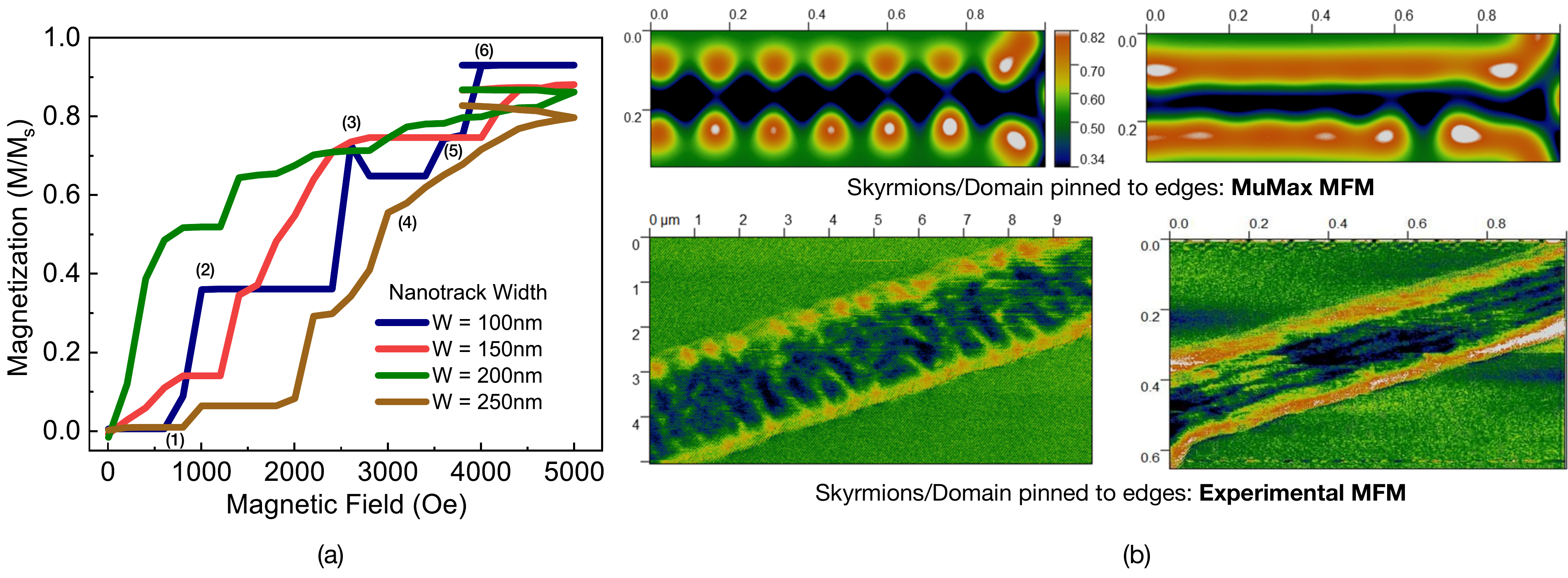}
    \caption{(a)~Discrete magnetization switching of the crossbar of different widths. (b)~MFM imaging shows DW - pinning to edges.}
    \label{fig:mag1}
\end{figure}

The fabricated crossbar spintronic devices based on the magnetic heterostructure are shown in Fig.~\ref{fig:Fig1}(a). The device consists of the $\mathrm{[Ta (2)/CoFeB (0.9)/MgO (2)]}_{20}$ based multilayer magnetic hall bar devices patterned on $\mathrm{Si/SiO_2}$ with gold contacts. The area of these fabricated devices is in the range (2 um$\times$50 um to 3 um$\times$50 um). The contacts are put on the lateral sides of the magnetic crossbar to provide the voltage from the pulse generator into one of the crossbar legs. While the resistance is measured across the transverse leg by the standard PPMS set-up and the Oscilloscope. Considering the 2 nm thick MgO dielectric which allows a very small tunneling current. The multilayer device structure can be considered as 20 magnetization-dependent resistors in parallel as shown by the equivalent resistor model in Fig.~\ref{fig:Fig1}(b). The magnetization of the magnetic layers is changed by the external magnetic ranging from (-650 mT to 650 mT). Fig.~\ref{fig:mag1}(b) shows the MFM image of the device showing the unperturbed multi-domain magnetic texture with stripe domains. Since the width of the Hall bar is 2 um and based on the thin film magnetic stack, we expect stabilization of the N\'{e}el-type domain wall having a width around 20 nm as revealed by micromagnetic simulations [Fig.~\ref{fig:Fig1c}]. The Hall measurements are performed using the standard lock-in technique. As shown in Fig.~\ref{fig:Fig2}(a) we observe discrete anomalous Hall resistance. For samples at room temperature, the magnetic anisotropy and saturation magnetization are small thus all magnetic layers switch at lower fields. At 300K the magnetization switches at 21 mT for the positive magnetic fields and 19 mT for the negative magnetic field. \\
Likewise, the switching field keeps on increasing from 21 mT to for lower temperatures and switching becomes more gradual with each resistance state separated by discrete steps. The discrete behavior and gradual switching are attributed to the fact that on lowering temperature the saturation magnetization and anisotropy start increasing. This increases the stary field effects and the domain wall pinning to the edges.  The thermal effects are high which reduces the domain wall pinning effect at this temperature. We observe discrete resistance states as all magnetic layers switch simultaneously at a 40 mT magnetic field. As we lower the temperature the anisotropy and saturation magnetization increase which increases the switching field as well as the stary field effect on different layers. Moreover, the thermally activated depinning is lowered resulting in the emergence of more discrete states as shown in Fig.~\ref{fig:Fig2}. For the temperature $T=120$ K we obtained 11 discrete resistance states as shown in Fig.~\ref{fig:Fig2}(b). The ON/OFF ratio or memory window obtained of the devices is around 66 which is quite better compared to other spintronic devices where ON/OFF is around 3. Furthermore, the achieved ON/OFF ratio provides a reliable synaptic application of the device.\\
In Fig.~\ref{fig:Fig2}(b) we show the obtained resistance states at different temperatures, at 300K only 2 states are observed, at 250K 3 resistance states, at 4 states at 200K, 8 at 150K, and 11 states at 120K. As we lower the temperature below 120K the number of resistance states starts decreasing. We measure 8 states for 80K, then again 11 states for 20K and 6 states for 2K depicted in Fig.~\ref{fig:Fig2}(c). This behavior can be explained by the fact that although by lowering the temperatures lowers the thermal effects and increases saturation magnetization, which results in discrete switching due to pinning/depinning effects added to individual layers switching independently. But anisotropy is also increasing thus on one hand the number of states should increase but increased anisotropy stabilizes the multilayer structure thus magnetic layers start switching unanimously which results in a reduced number of states as observed. In Fig.~\ref{fig:Fig2}(d) the temperature dependence of RXX and RXY is shown, the approximate $\mathrm{\sigma_{xx}\cong 4.22  ~k\Omega cm^{-1}}$  indicating a bad metal behavior. We observe overall discrete and sharp increase of RXY with increasing RXX, which is expected. But the discrete jumps indicate the increasing stray field effects at lower temperatures on the different ferromagnetic layers. Thus, pinning/deppining is increasing with lowering temperatures.

We furthermore simulated similar and scaled crossbars in the micromagnetic software MuMax \cite{30,31}. The simulation parameters such as saturation magnetization and anisotropy were taken from the experimental results (VSM). The parameters and simulation details are given in the 
In simulations, we considered crossbar nanotrack width from 100 nm to 300 nm in intervals of 50nm. When perturbed by the external magnetic field ranging from (-5000 Oe to 5000 Oe) we observe discrete magnetization switching for all the nanotrack widths. The discrete magnetization behavior is most dominant for the lowest width $W=100$~nm. For the nanotrack length $L= 1$~um, 5 stable discrete magnetization states were observed as shown in Fig.~\ref{fig:mag1}(a). As the width increases the magnetization discrete switching behavior starts fading. Although 4 discrete states are observed for $W=250$ nm, the switching field gap between the states is reduced. In 250 nm the magnetization switching is gradual whereas in 100 nm nanotrack, the magnetization switches in sharp steps as shown in Fig.~\ref{fig:mag1}(a). 
\begin{figure}[htbp]
  \centering
  \includegraphics[width=\textwidth]{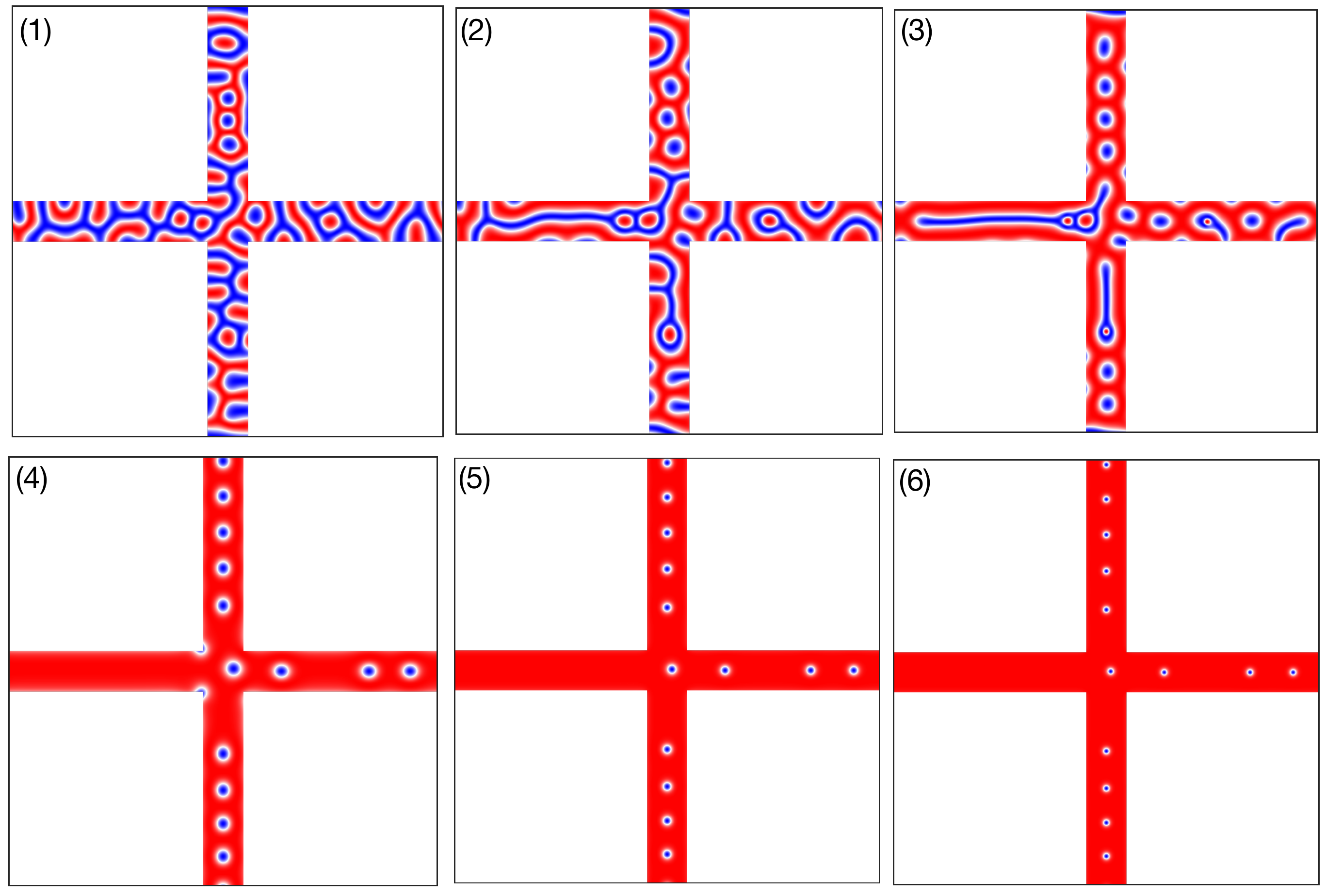}
  \caption{Repeatability of the discrete magnetization behavior with the application of 1 ns magnetic field pulses.}
  \label{fig:mag2}
  
  \vspace{7mm}
  
  \includegraphics[width=\textwidth]{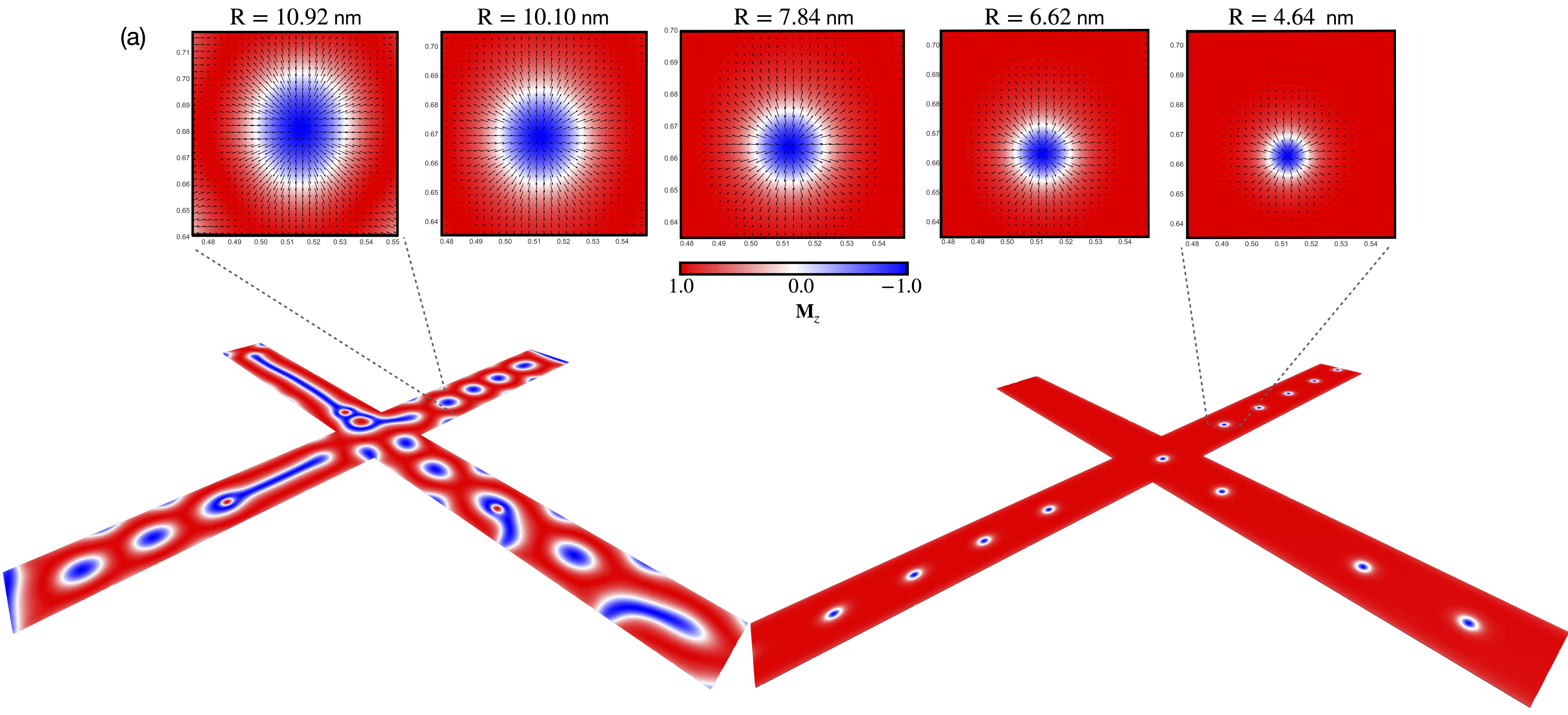}
  \caption{Magnetization texture through the hysteresis loop shows switching occurring via stripe domain conversion into skyrmions.}
  \label{fig:mag3}
\end{figure}

In corresponding magnetization textures obtained from the simulations, we observe switching from stipe domains to stable skyrmions as the field changes. Fig.~\ref{fig:mag2} shows the magnetization profile of the single FM layer at the different switching points. At zero magnetic field the stripe domains are pinned to the track edges, With the application of the external magnetic field in the Z-axis we observe the depinning and stabilization of the skyrmions taking place. The numbers [(1) to (6)] show the discrete switching points as shown in Fig.~\ref{fig:mag1} and \ref{fig:mag2}. For cases (1), (2) and (3) we see normalized magnetization switching from 0 to 0.7 while from (4) to (6) the normalized magnetization reaches a value of 0.9. Corresponding to switching points (1), (2) and (3) we observe the depinning and conversion of stripe domains into the skyrmions. This accounts for the maximum change (about 77\%) in the magnetization of the device. After switching point (3) the skyrmions are stable and the external field reduces the size of the skyrmions as seen in the corresponding texture in Fig.~\ref{fig:mag2} [(4), (5) and (6)] and in Fig.~\ref{fig:mag3}. This increases the magnetization more linearly and overall change is about 23\%. The skyrmion radius calculated using nearest linear interpolation reduces gradually from $R = 10.92$ nm to $R = 4.64$ nm. \\
The magnetic field ranging from (-5000 Oe to 5000 Oe) is applied perpendicular to the device. In the positive field regime, the N\'{e}el skyrmions with polarity $+1$ are stabilized from the stripe domain phase. While reducing the field from high to low, skyrmioniums are stabilized in the Hall bar which stabilizes into skyrmion polarity $-1$ as we increase the field in the -z-direction. Depending upon the direction of the magnetic field the skyrmions change the polarity from $+1$ to $-1$ and vice-versa (see SV1 and Fig.~\ref{fig:mag2} and ~\ref{fig:mag3}). The experimental MFM images of the fabricated device and the micromagnetic MFM of the scaled device are shown in Fig.~\ref{fig:mag1}(b) (bottom). In both cases, we observe clear skyrmion/domains pinning to the edges of the device.  

\section{Multilayer Spintronic Synapse }
\begin{figure}[!h]
    \centering
    \includegraphics[width = \linewidth]{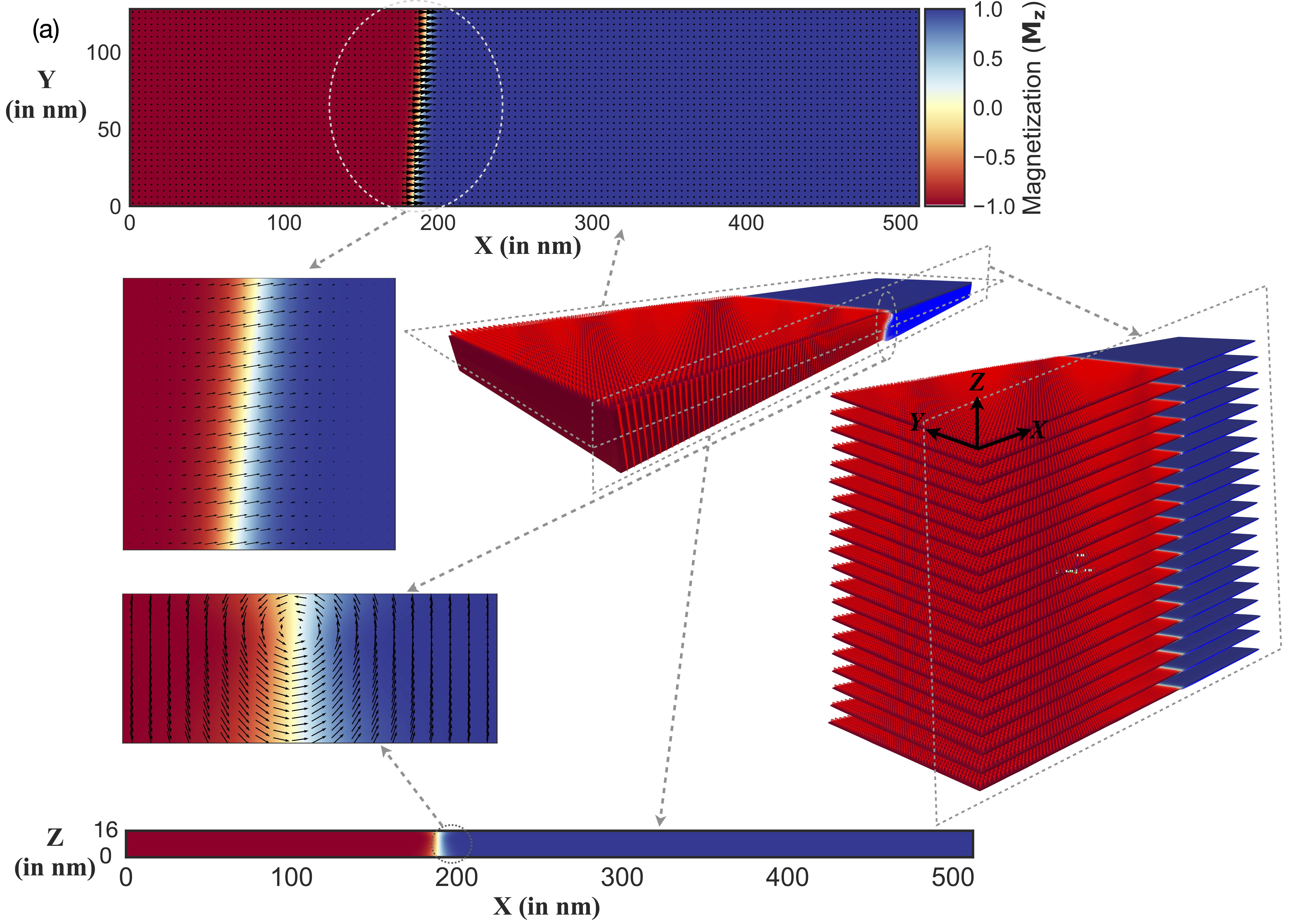}
    \caption{(a) Micromagnetic texture of the 20-Layer stack.}
    \label{fig:Fig1c}
\end{figure}
\begin{figure}[ht]
    \centering
    \includegraphics[width = \linewidth]{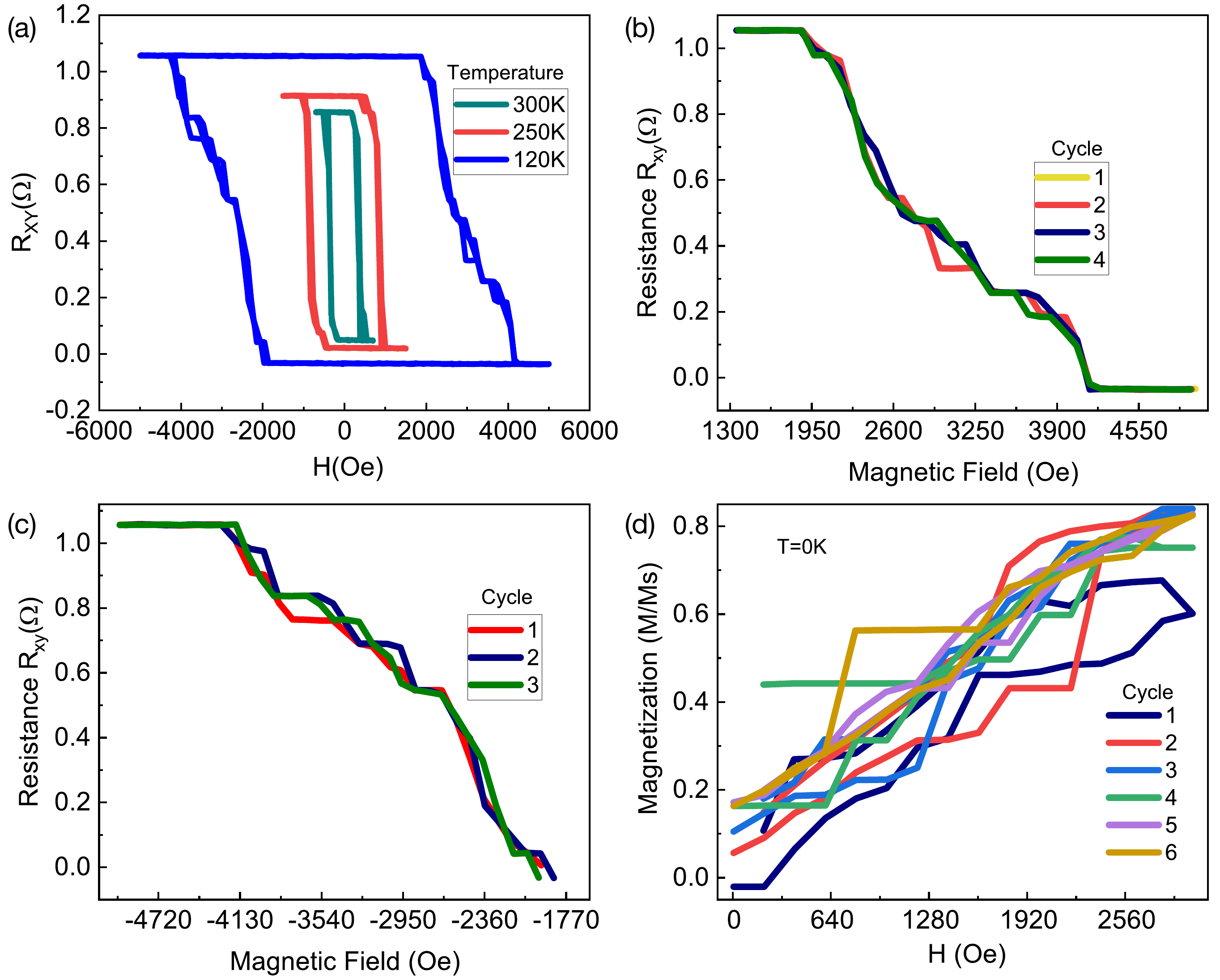}
    \caption{(a) Anomalous Hall resistance for T=300K, 250K and 120K (repeated 3 times). (b-c) Discrete resistance at positive and negative field range (T=120K), and (d) Micromagnetic simulation showing magnetization evolution shows discrete behavior repeated 6X.}
    \label{fig:Fig4}
\end{figure}

We repeat the hysteresis simulation to check the feasibility of using these multilayer spintronic devices as the synapses. As shown in Fig.~\ref{fig:Fig4}(a) the measurement was repeated three times for the different temperatures to check the repeatability of the discrete resistance behaviour shown by the crossbar. We observe the pinning/depinning behaviour thus, discrete resistance states as shown in Fig.~\ref{fig:Fig4}(a) for the temperatures 300K, 250K and 120K. Apart from a minor deviation in the value of depinning fields, the discrete states are observed during all the measurement cycles. For temperature $T=120$K, Fig.~\ref{fig:Fig4}(b-c) show that when properly tuned by the external magnetic field we can increase or decrease the resistance in a discrete fashion. This opens the possibility of employing this device as multibit memory or synapse for neuromorphic computing applications. To ascertain it we simulated the repeatability measurement for nano-track width $W = 100$ nm in MuMax as shown in Fig.~\ref{fig:Fig4}(d). The measurement was repeated 6 times and as shown in Fig.~\ref{fig:Fig4}(d), when the magnetic field is limited in the range of (0 to 3000Oe the discrete behaviour is observed throughout 6 cycles.
The devices show hysteresis indicating a memory effect, but we observe some randomness in the hysteresis cycles as shown in Fig.~\ref{fig:Fig4}(a) [cycle 1 to 6].  

The repeatability of the discrete resistance indicates the potential application of the device as a synapse in neural networks. Where the probability of the stochastic behaviour is controlled by the (a) temperature (b) pinning distribution and (c) external field/current. We also show the evolution of skyrmion topological charge during the different phases. For the first two cycles of pulse number (50), the topological charge increases from 0 to $-28$, after that skyrmion density varies between ($-28$ and $-20$). This shows the magnetization contribution from the skyrmions during the potentiation/depression. We observe the conversion of stripe domains into skyrmions throughout the cycles as shown in [SV3]. 

We further simulated the current response of the multilayer devices with 15 layers and 20 layers. The device is pulsed with 1 ns pulse of current density $5\times10^{11}$~$\mathrm{A/m^2}$ after every 1.5 ns. We clearly observe the synaptic potentiation/depression emulated by the device in terms of magnetization profile. The resistance of the device depends upon the magnetization profile as,
\begin{equation}
\rho_{x y}=\rho^{O}+\rho^{A}+\rho^{T}=R_{O} B+R_{S} \mu_{0} M_{Z}+R_{O} P B_{e m}^{z}
\label{eq1}
\end{equation}
Where $\rho^{0}$ is the ordinary Hall resistivity which has negligible contribution in the magnetic materials, and second is the anomalous Hall resistivity. The third term represents the topological Hall resistivity added by the skyrmions and can be considered to have a very small contribution.\\
The emergent magnetic field is given by \cite{34}
\begin{equation}
    B_{Z}^{e}=\frac{\Phi_{Z}^{e}}{A}=-\frac{h}{e A} \iint \frac{1}{4 \pi} \boldsymbol{m} \cdot\left(\frac{\partial \boldsymbol{m}}{\partial x} \times \frac{\partial \boldsymbol{m}}{\partial y}\right) d x d y
\end{equation}
Leading to topological resistivity
\begin{equation}
    \rho_{x y}^{T}=P R_{o}\left|\frac{h}{e}\right| \frac{1}{A}
\end{equation}
The other method of reading the device is by tunnel magnetoresistance TMR. Depending upon the magnetization profile the resistance of the MTJ is given by \cite{35},
\begin{equation}
R_{\text{MTJ}}=R_{\text{AP}} \frac{\left[1-\hat{m} \cdot \hat{m}_{P}\right]}{2}+R_{\text{P}} \frac{\left[1+\hat{m} \cdot \hat{m}_{P}\right]}{2}
\label{eq4}
\end{equation}
Where $R_{\text{AP}}/R_{\text{P}}$ are the antiparallel/parallel resistances and $\hat{m}/{\hat{m}}_P$ is the free layer/pinned layer normalized magnetization.\\

\begin{figure}[t]
    \centering
    \includegraphics[width = \linewidth]{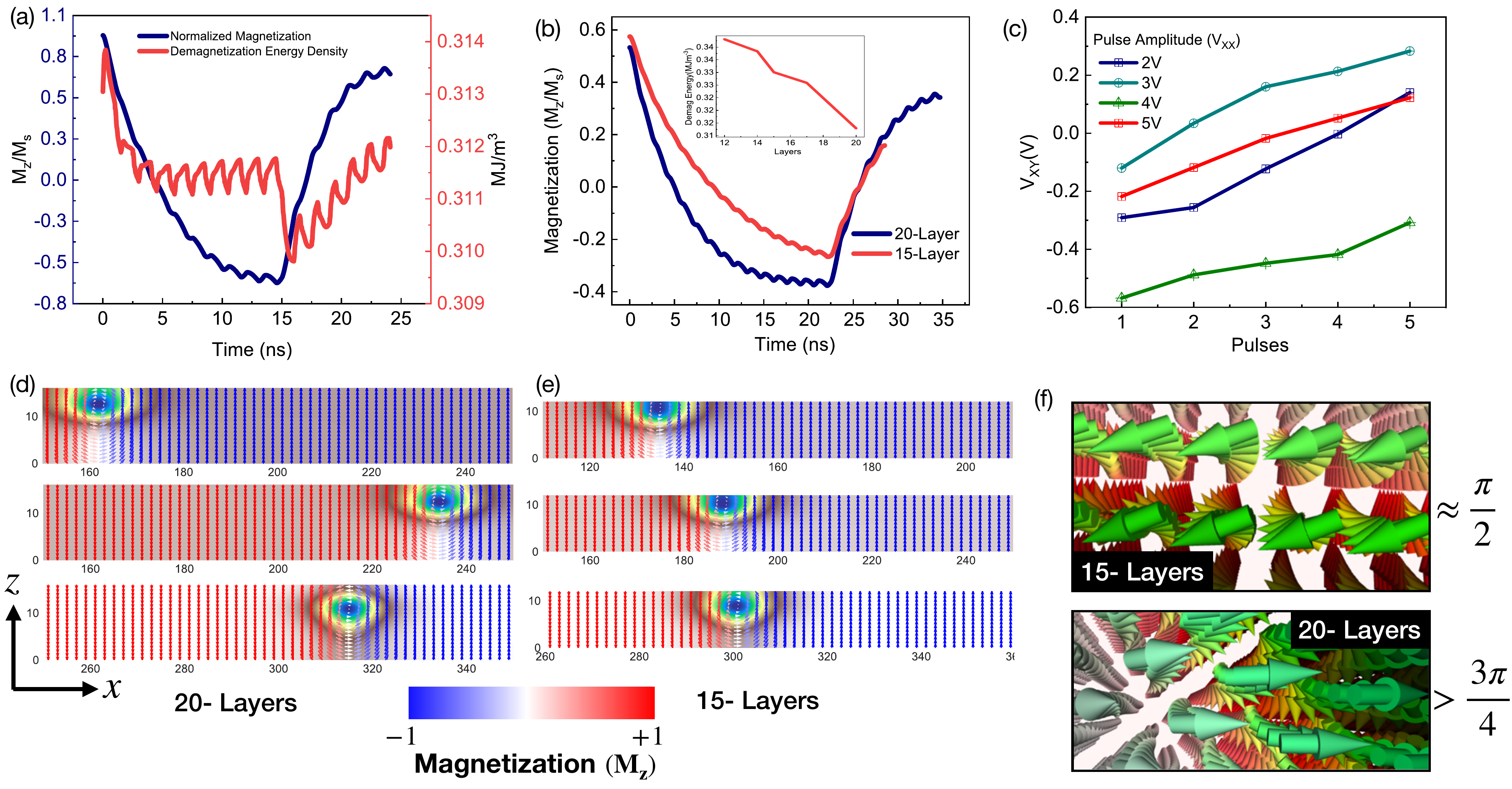}
    \caption{(a) Normalized magnetization evolution and associated demagnetization energy: explaining the behavior of DW motion. (b)  SOT controlled DW-Synapse: Magnetization profile (potentiation/depression) in 15-layer and 20-Layer device. (micromagnetic). (c) DW shape during negative and positive current explaining the sudden drop in demagnetization energy on reversal of current. (d-e) (f) Current controlled synapse (measured). }
    \label{fig:Fig5}
\end{figure}

\begin{figure}[t]
    \centering
    \includegraphics[width = \linewidth]{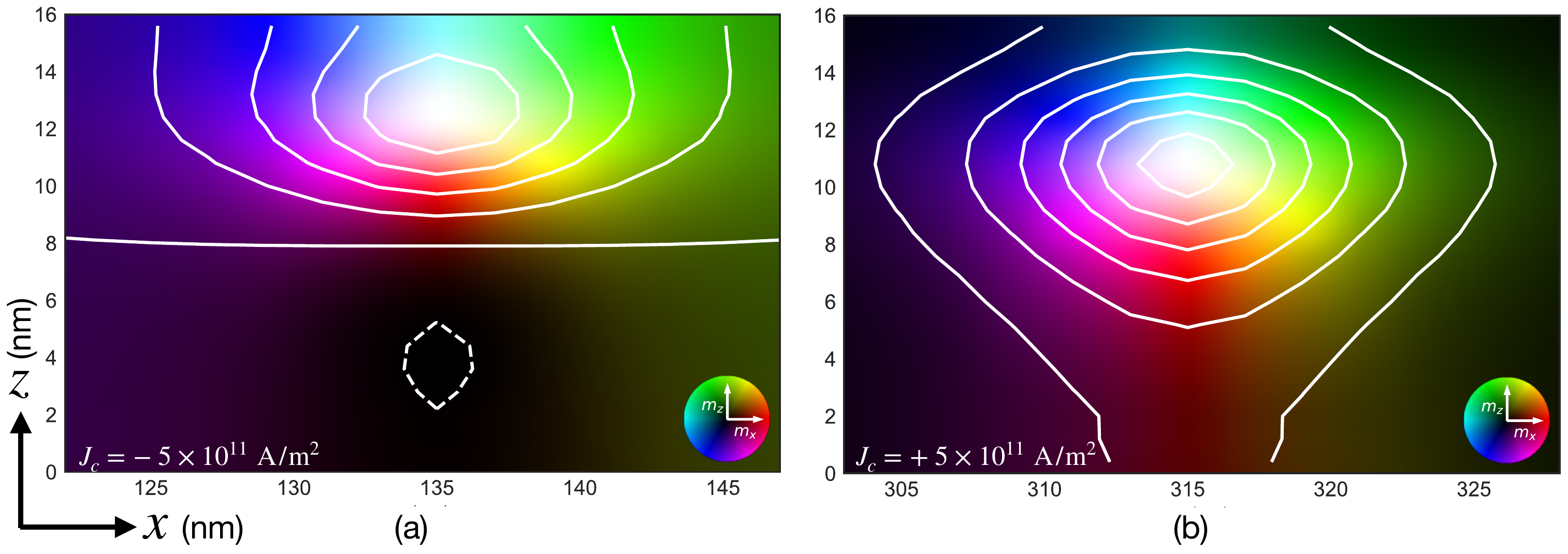}
    \caption{Contour depiction of the magnetization vector within the magnified cross-sectional region of (a) 10 layers (b) 20 layers.}
    \label{fig:FigX}
\end{figure}

\begin{figure}[t]
    \centering
    \includegraphics[width = \linewidth]{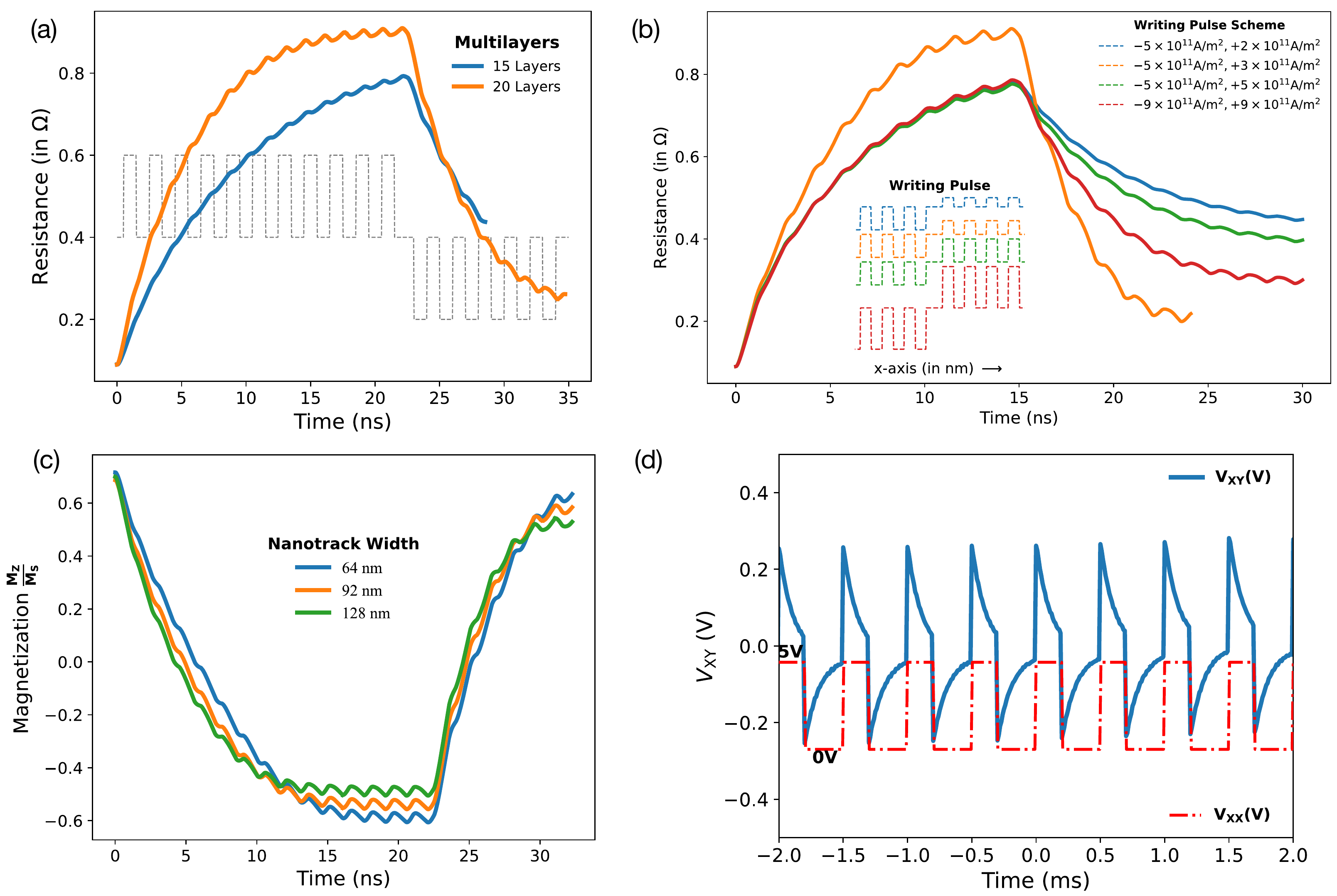}
    \caption{(a) SOT controlled Multilayer DW-Synapse: Conductance (potentiation/depression) in 15 Layer and 20 Layer device. (micromagnetic) (b) Writing pulse scheme for synaptic linearity improvement. (c) Magnetization evolution for different nano-track widths. (d) }
    \label{fig:Fig6}
\end{figure}

 In micromagnetic simulations the current pulses with amplitude $J_c=-5\times{10}^{11}~\mathrm{Am^{-2}}$, pushes the domain wall in the $+x$ direction whereas the DW motion direction is reversed for $J_c=+5\times{10}^{11}~\mathrm{Am^{-2}}$. Depending upon the DW position the net magnetization of the device varies which leads to the variation in the resistance of the device. Depending upon the reading mechanism [AHE – Eqn. \ref{eq1}] We compute the resistance by mapping magnetization to the measured resistance. We further consider the [TMR– Eqn. \ref{eq4}] reading to compute the synapse conductance evolution. The magnetization potentiation/depression of the 15-layer and 20-layer synaptic device, with current pulses is shown in Fig. \ref{fig:Fig5}(a-b).
 When the DW starts moving away from the left edge, at the beginning we observe fast magnetization evolution because DW is going away from the edge which results in reduced demagnetization energy density, thus increased DW velocity. As the DW reaches the center reaches a minimum and with more current pulses Demagnetization energy starts gradually increasing until we reverse the current, this results in a sudden drop in the demagnetization energy due formation of a full Bloch DW in the perpendicular direction as shown in Fig. \ref{fig:Fig5}(d-e). On the application of $J_c=-5\times{10}^{11}~\mathrm{Am^{-2}}$ the demagnetization energy is minimized by the twist in the perpendicular axis, we observe the bottom 10 layers to be stable whereas the top 10 layers form an incomplete Bloch DW. For $J_c=+5\times{10}^{11}~\mathrm{Am^{-2}}$  the DW gains full rotation, and more layers flip such that the first 6-7 bottom layers are stable and the remaining 13 layers form a full Bloch DW, as clearly depicted in the contour plot of the cross-section in Fig.~\ref{fig:FigX}. \\
Thus, resulting in reduced Demag energy, hence increased DW velocity. Furthermore, for the 20-layer device, we observe fast switching which indicates increased DW velocity, in comparison to the 15-layer  device [see Fig. \ref{fig:Fig5}(b)]. This phenomenon is explained by the inset Fig. \ref{fig:Fig5}(b), interestingly, we observe the lowering of the demagnetization energy with the increasing number of FM layers. due to the increased number of layers, the spins vertically rotate almost by $\pi$ thus, forming the Bloch domain wall which reduces the demagnetization energy. Compared to the case of 15 layers the rotation is by about $\frac{\pi}{2}$ resulting in higher demagnetization energy.
Fig. \ref{fig:Fig5}(c) shows the measured current-controlled device operation, here we applied the 200 us voltage pulses with amplitude 5V and time period 500 us across the length of the device. The signal generated across the transverse arm is collected by the oscilloscope. We observe potentiation of the transverse voltage which we attribute to the change in the magnetization due to the SOT and STT. The anomalous Hall resistance and planner Hall resistance depend on the net magnetic moment of the device. As shown in Fig. Fig. \ref{fig:Fig5}(c) and Fig. \ref{fig:Fig6}(d) the resistance change leads to the variation in the voltage drop across the device. We observed an increasing voltage drop with respect to the number of current pulses (200us) being applied across the X-axis. The observed resistance is the combined effect of AHE (lateral), planar Hall effect PHE (lateral) and tunnel magnetoresistance (TMR) measured vertically across the device as shown in Fig. \ref{fig:Fig1}(b) equivalent resistance model.
Fig.~\ref{fig:Fig6}(a) shows the synaptic conductance potentiation/depression with the application of the current pulses in 15-layer and 20-layer devices. As discussed above in comparison to the 15-layer device, the conductance evolves faster for the 20-layer device due to reduced demagnetization energy. For the same writing current density, and time, the performance of a 20-layer device is better. Thus, demagnetization energy can be utilized for reducing the writing energy dissipation in these multilayer devices. As seen in Fig.~\ref{fig:Fig6}(a) the conductance/resistance switches sharply at the beginning of the positive current due to reduced demagnetization energy density, followed by the gradual switching as DW approaches the edges. This leads to the increased non-linearity NL of the synapse. We compute the NL by using the following methodology,
\begin{equation}
G_{\text{LTP}}\left(V_{g}, V_{d}\right)=\beta\left[1-e^{-(P / \alpha)}\right]+G_{\min }\left(V_{g}, V_{d}\right)
\end{equation}
\begin{equation}
G_{\text{LTD}}\left(V_{g}, V_{d}\right)=\beta\left[1-e^{-\left(\frac{P-P_{M}}{\alpha}\right)}\right]+G_{\max }\left(V_{g}, V_{d}\right)
\end{equation}
\begin{equation}
\beta=\frac{\left[G_{\max }\left(V_{g}, V_{d}\right)-G_{\min }\left(V_{g}, V_{d}\right)\right]}{\left(1-e^{-\frac{P_{M}}{\alpha}}\right)}
\end{equation}
 Where, $G_{\text{LTP}}$ is the synaptic conductance, $G_{\text{min}}$ is the minimum conductance and $G_{\text{max}}$ is the maximum conductance achieved by the device, $\alpha$ is the non-linearity fitting parameter, $P_M$ is the maximum pulse number, $\beta$ is the function of $G_{\text{min}}$,  $G_{\text{max}}$, $\alpha$ and $P_M$.  To improve the linearity of the device, we propose varying pulse schemes as shown in Fig.~\ref{fig:Fig6}(b). The current density is reduced during the depression cycle, for $J_C=+2\times{10}^{11}~\mathrm{Am^{-2}}$ the non-linearity (NL) $\alpha= -3.19$, for  $J_C=+3\times{10}^{11}~\mathrm{Am^{-2}}$ (NL)$ \alpha= -3.36$, for $J_C=+5\times{10}^{11}~\mathrm{Am^{-2}}$ (NL) $\alpha= -3.8$ and for $J_C=\mp9\times{10}^{11}~\mathrm{Am^{-2}}$ (NL) $\alpha= +2.29/-3.19$. Thus, we can reduce the NL by either reducing the write current density during the resistance depression or increasing the current density to a much higher value where it dominates the demagnetization effect as seen for $J_C=\mp9\times{10}^{11}~\mathrm{Am^{-2}}$. Fig.~\ref{fig:Fig6}(c) shows the scaling of the synapse behavior with the scaling nanotrack width, the magnetization potentiation/depression for = 64 nm device shows better linearity compared to the 128 nm devices due to reduced demagnetization energy. The measured potentiation of multilayer devices is shown in Fig. 9(d). The application of a 5V (200us) voltage pulse results in the voltage drop $V_{XY}$ across the transverse arm (short-term potentiation STP). The voltage drop relaxes back to 0V in the absence of the $V_{XX}$ pulse. As the device is stressed with more voltage pulses we observe gradual long-term potentiation LTP. This phenomenon is linked to the magnetization switching and relaxation in the device. The pinning of DWs during the relaxation phase can account for the long-term potentiation. These measured and simulated results clearly promise the potential of multilayer devices-based neuromorphic computing.  

\section{Multilayer Spintronics Synapse ANN for MNIST Data Recognition}
\begin{figure}[htbp]
  \centering
  \includegraphics[width=\textwidth]{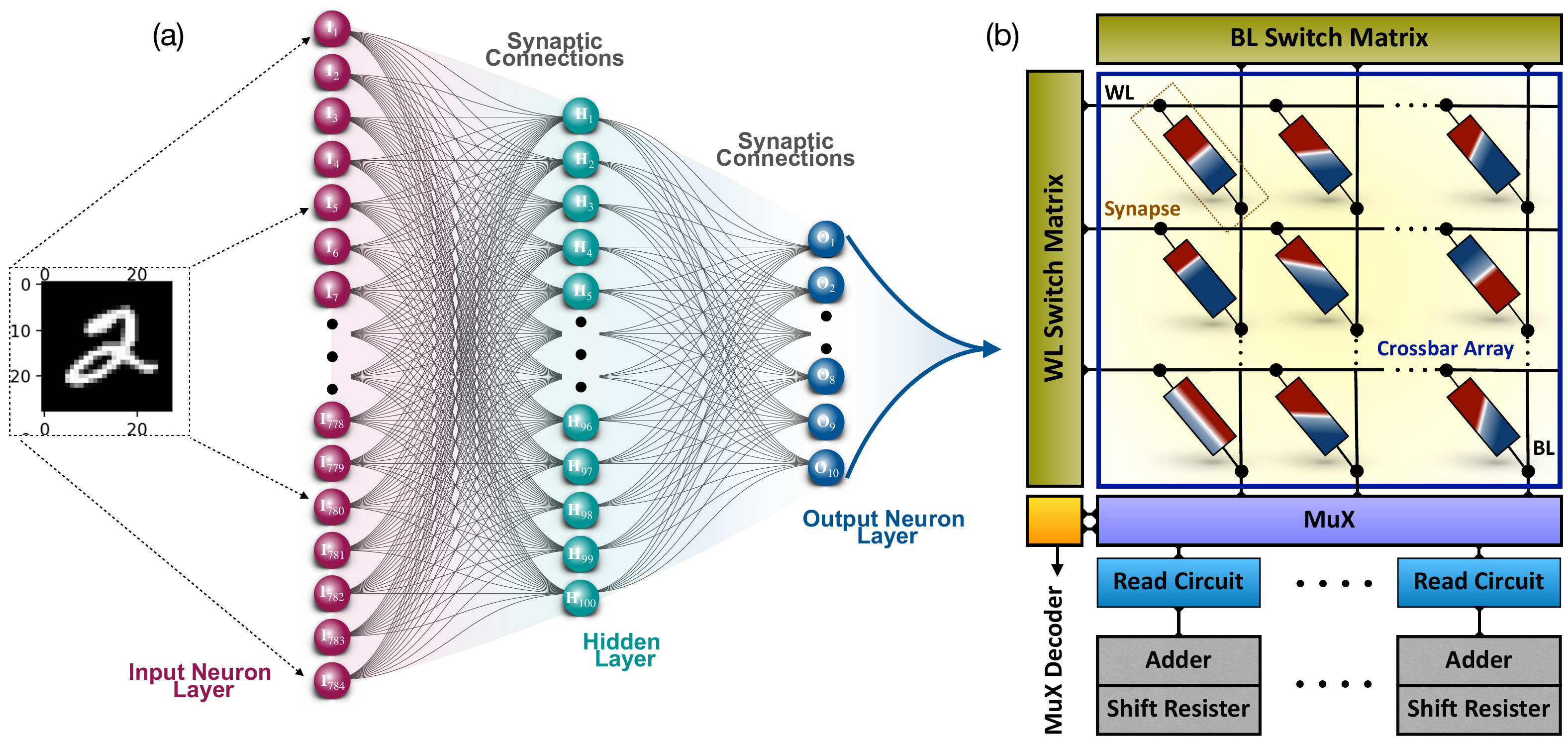}
  \caption{Illustration of the 3-Layer FCNN architecture for MNIST data recognition. Recognition accuracy increases throughout 125 epoch training for different write pulse schemes and the ideal case.}
  \label{fig:Fig10}
  
  \vspace{7mm}
  
  \includegraphics[width=\textwidth]{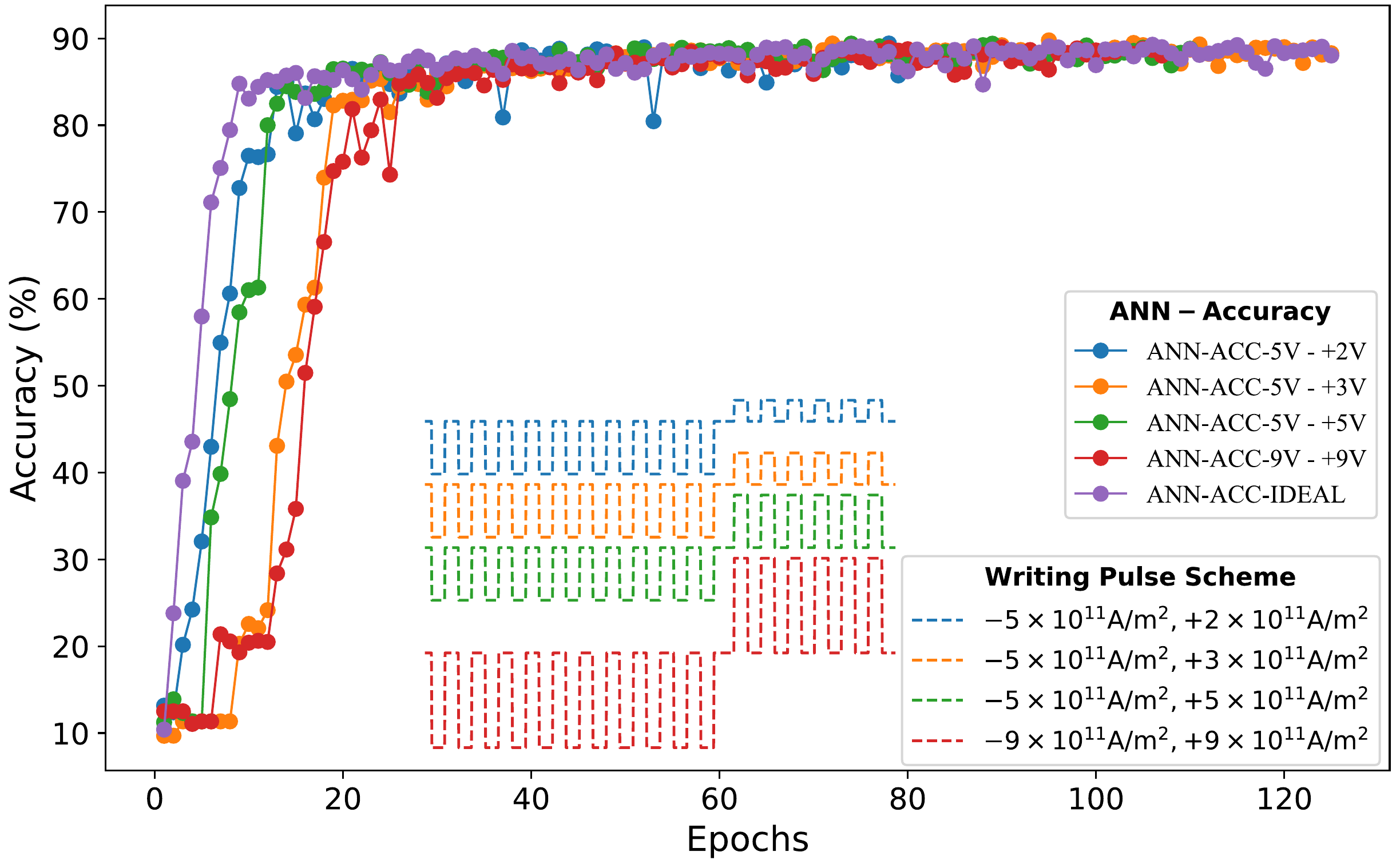}
  \caption{The recognition accuracy is above 89\% for all cases.}
  \label{fig:Fig11}
\end{figure}

The performance of FM multilayer synapse is evaluated using the MLPNeuroSim \cite{33} integrated framework for on-chip training and inference. The simulator is used to provide the benchmark from the device to the circuit level. We considered a simple 3-layer feedforward neural network FNN having 784 input neurons, 100 hidden neurons and 10 output neurons as shown in Fig.~\ref{fig:Fig10}(a). The architecture is trained and tested for the classification of the handwritten MNIST digits using the stochastic gradient descent algorithm SGD. The input to the network is the pixel value from the $28\times28$ matrix representing the handwritten digit. The synaptic conductance values act as the weight of the neural network. Here the weights are mapped to the conductance values of the devices. Fig.~\ref{fig:Fig10}(b) shows the crossbar matrix of the DW devices and the peripheral circuitry such as the read circuit, adder and shift registers.

As shown in Fig.~\ref{fig:Fig11} the classification accuracy of the devices with different write pulse schemes indicates that the proposed devices can learn and recognize the MNIST dataset with up to 90\% accuracy. Which is very well in range for acceptable recognition performance benchmarked for other memristors and ideal software-based 3-layer FCNN. The system level read energy and write energy is $1.47\times 10^{-4}$ J and $4.8\times 10^{-3}$ J. The read latency and write latency are $1.98\times10^{-2}$ s and $2.38\times 10^{-1}$ s. These results provide a clear benchmark for the realization of hardware neural accelerators based on the DW-MTJ devices. 

\section{Multilayer Spintronics Devices as Cryogenic Memory for Scalable Quantum Computing}
\begin{figure}[htbp]
    \centering
    \includegraphics[width = \linewidth]{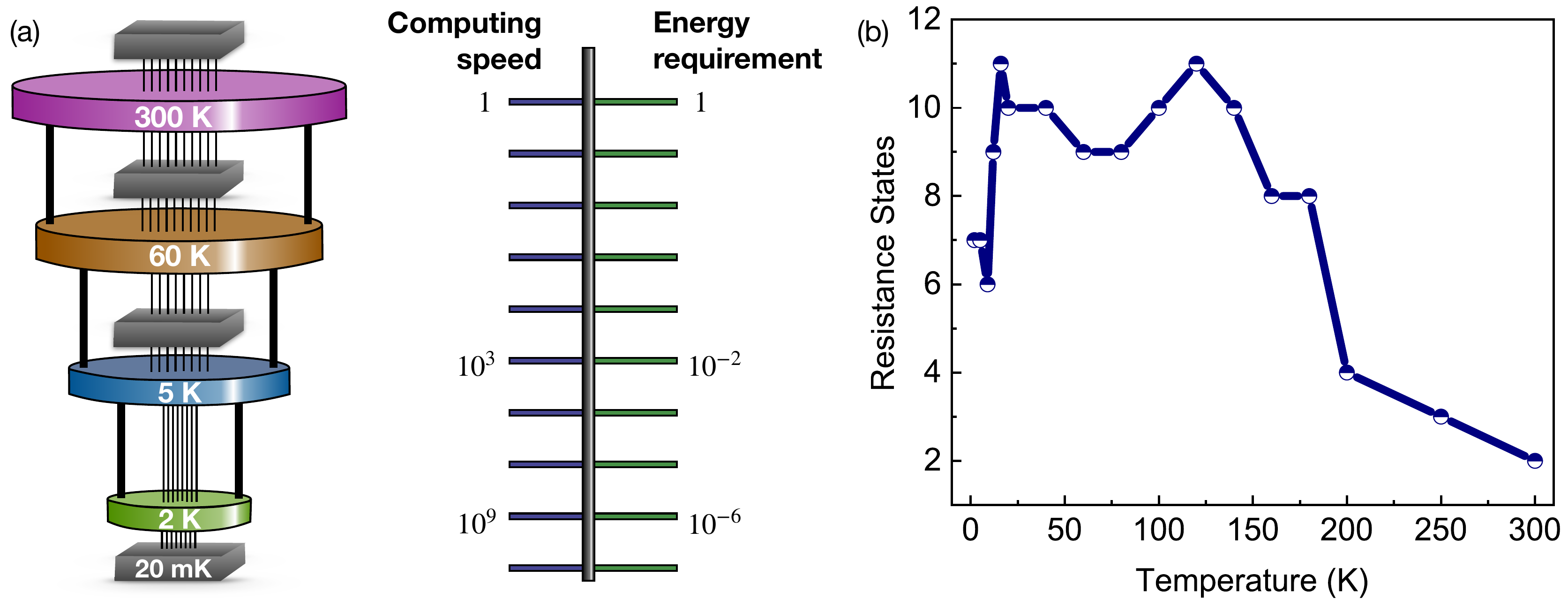}
    \caption{(a) Proposed application of the device as the cryogenic memory in the realization of a super-conducting control processor for QC (B). (b) Discrete resistance states at different temperatures.}
    \label{fig:Fig12}
\end{figure}

Although Quantum Computers (QC) have seen a huge leap in development recently, the realization of a scale and compact QC is still a Challenge. The typical QC system has three major sub-systems Quantum substrate (q-bits), the control processor and a memory block. Normally the q-bits are placed at a few mT temperatures while the control processor is placed at room temperature (300K). The two blocks are connected by the long control cables which work fine for a few q-bit systems. But this system faces scaling issues as for an increased number of q-bits the control cables have to increase. To overcome this issue the best option that has been explored is to keep the control processor (superconducting) at 4K itself as shown in Fig. \ref{fig:Fig12}(a) \cite{36}. This scheme further demands the cryogenic memory to reduce thermal leakage. The other option is to consider a hierarchical system as shown in Fig. \ref{fig:Fig12}(a), which shows cryogenic memories at different temperatures. Considering the discrete resistance states sown by the fabricated device We propose an interesting idea where the multilayer spintronic devices having discrete resistance states could be used as the cryogenic memory for scalable QC systems of the future. Interestingly, we have shown a multilayer spintronic device exhibiting 6 states at 2K, 11 states at 60K, 11 states at 120K and finally 2 states at 300K.
So, at low temperatures, the fabricated device can be used to store multiple control bits on a single device which is highly desirable considering the cooling system requirements needed for these memories. Moreover, the same device as memory across the ladder should reduce the issues of integration. We can anticipate a future hybrid computing system that involves a QC, cryogenic neuromorphic systems based on these memories and a CMOS system.

\subsection{Fabrication and Characterization}
The Rotaris magnetron sputtering system deposited the magnetic thin films at room temperature on a 4-inch thermally oxidized Si wafer. A 300 nm bottom layer of $\mathrm{SiO_2}$ is deposited prior to stack deposition by using thermal oxidization. Then, the multilayer stack is grown using ultrahigh vacuum magnetron sputtering at room temperature in $5 \times 10^{-8}$ mbar.
The multilayer structure consists of, from the substrate side, a 300 nm $\mathrm{SiO_2}$ insulating layer.  Finally, a stack of $\mathrm{SiO_2(300nm)/[Ta (5~{nm})/CoFeB(x)/MgO (2~{nm})]\times20}$ is formed, where MgO layers were deposited by RF magnetron sputtering, and the other layers were deposited by DC magnetron sputtering.
After the film deposition, we spin-coated AZ5214 photoresist with a thickness of 1.6 um and performed a hard bake for 2 mins at $110^{\circ}$C for positive tone use. We then patterned the crossbars on the resist using conventional photolithography. Ion beam etching using Ar gas was used to remove the exposed magnetic stacks outside the resist mask. During the etching, we monitored the conductivity of the etched region and stopped the etching when the signal from the 300-nm-thick silicon oxide layer appeared. To form electrical contacts, Ti (10 nm)/Au (100 nm) was deposited on the sides of the crossbars through sputtering at a rate of 0.8 nm/s. The electrical contact pads were defined using photolithography and lifted off by immersion in acetone with ultrasonic processing for 5 mins.

\subsection{Characterization and Imaging}
The magnetic characterization of the samples for thickness optimization was done using normal vibration sample magnetometry (VSM) at room temperature. After VSM we performed the imaging of the samples with multi-domain magnetic characteristics using magnetic force microscopy (MFM) based- Dimension Icon SPM.  For probing we used CoIr coated MFM tip provided by Bruker Inc.  The Hall measurements were done using a standard Hall measurement system Quantum Design PPMS capable of applying dc magnetic fields. We measured the crossbar devices at temperatures ranging from (2K to 300K) and at different magnetic fields depending upon the temperature.  Lastly, a signal generator was used to supply voltage pulses through the MTJ, and an oscilloscope was used to measure the output voltage response of the devices.

\section{Micromagnetics}
Magnetic skyrmions are described using their topological or skyrmion number $Q$, calculated as follows \cite{77}
\begin{equation}
Q=\frac{1}{4 \pi} \iint \boldsymbol{m} \cdot\left(\frac{\partial m}{\partial x} \times \frac{\partial \boldsymbol{m}}{\partial y}\right) d x d y
\end{equation}
The spins projected on the XY-plane and normalized magnetization vector $\boldsymbol{m}$ can be determined by the radial function $\theta$, vorticity $Q_v$\ and helicity $Q_h$:
\begin{equation}
m(r)=\left[\sin (\theta) \cos \left(Q_{v} \varphi+Q_{h}\right), \sin (\theta) \sin \left(Q_{v} \varphi+Q_{h}\right), \cos (\theta)\right]
\end{equation}
The vorticity number is related to the skyrmion number as follows: 
\begin{equation}
Q=\frac{Q_{v}}{2}\left[\lim _{r \rightarrow \infty} \cos (\theta(r))-\cos (\theta(0))\right]
\end{equation}
Micromagnetic simulations were performed using MuMax having the Landau–Lipschitz–Gilbert (LLG) equation as the basic magnetization dynamics computing unit. The LLG equation describes the magnetization evolution as follows:
\begin{equation}
\frac{d \hat{m}}{d t}=\frac{-\gamma}{1+\alpha^{2}}\left[\boldsymbol{m} \times \boldsymbol{H}_{\text{eff}}+\boldsymbol{m} \times\left(\boldsymbol{m} \times \boldsymbol{H}_{\text{eff}}\right)\right]
\end{equation}
where $\boldsymbol{m}$ is the normalized magnetization vector, $\gamma$ is the gyromagnetic ratio, $\alpha$ is the Gilbert damping coefficient, and 
\begin{equation}
\boldsymbol{H}_{\text{eff}}=\frac{-1}{\mu_{0} M_{S}} \frac{\delta \boldsymbol{E}}{\delta \boldsymbol{m}}
\end{equation}
is the effective MF around which the magnetization process occurs. The total magnetic energy of the free layer includes exchange, Zeeman, uniaxial anisotropy, demagnetization, and DMI energies.
\begin{equation}
\boldsymbol{E}(\boldsymbol{m})=\int_{V}\left[A(\nabla \boldsymbol{m})^{2}-\mu_{0} \boldsymbol{m} \cdot H_{\text{ext}}-\frac{\mu_{0}}{2} \boldsymbol{m} \cdot H_{d}-K_{u}(\hat{u} \cdot \boldsymbol{m})+\varepsilon_{\text{DM}}\right] d v
\end{equation}
where $A$ is the exchange stiffness, $\mu_0$ is the permeability, $K_u$ is the anisotropy energy density, $H_d$ is the demagnetization field, and $H_{ext}$ is the external field; moreover, the DMI energy density is then computed as follows:
\begin{equation}
\boldsymbol{\varepsilon}_{\mathbf{DM}}=D\left[m_{z}(\nabla \cdot \boldsymbol{m})-(\boldsymbol{m} \cdot \nabla) \cdot \boldsymbol{m}\right]
\end{equation}
The spin–orbit torque is then added in the form of modified STT in MuMax. 
\begin{equation}
\boldsymbol{\tau}_{\mathbf{S O T}}=-\frac{\gamma}{1+\alpha^{2}} a_{J}[(1+\xi \alpha) \boldsymbol{m} \times(\boldsymbol{m} \times \boldsymbol{p})+(\xi-\alpha)(\boldsymbol{m} \times \boldsymbol{p})]
\end{equation}
$$
a_{J}=\left|\frac{\hbar}{2 M_{S} e \mu_{0}} \frac{\theta_{\text{SH}} j}{d}\right| \quad \text { and } \quad \boldsymbol{p}=\operatorname{sign}\left(\theta_{\text{SH}}\right) \boldsymbol{j} \times \boldsymbol{n}
$$
where $\theta_{\text{SH}}$ is the spin Hall coefficient of the material, $j$ is the current density, and $d$ is the free layer thickness. The resistance of the proposed skyrmion MTJ synapse is then computed using the compact model presented in the main discussion Eqn.~\ref{eq4}. We then consider the magnetization profile of the free layer and feed it to our model, which computes the resistance of the MTJ device as follows:
\begin{equation}
R_{\text{syn}}=\frac{V_{\text{syn}}}{I_{\text{syn}}}
\end{equation}
Table.~\ref{tab:table1}
\begin{table}[ht]
    \centering
    \resizebox{\textwidth}{!}{%
    \begin{tabular}{|c|c|c|c|c|c|c|c|c|}
    \hline
    \textbf{Grid Size} & \textbf{Cell Size} & \textbf{Anisotropy, $\mathcal{K}_u$} & \textbf{Saturation Mag,} $\mathrm{M}_s$ & \textbf{Exchange Stiffness, $\mathcal{A}$} & \textbf{DMI, $\mathcal{D}$}  & \textbf{IEC, $\mathcal{J}$} &$\alpha_{H}$ \\
    & (nm) & $~(\mathrm{J/m^2})$& $\mathrm{A/m}$ & (J/m) & $(\mathrm{J/m^2})$ & ($\mathrm{J/m^2}$)&  \\
    \hline
    256, 64, X\footnote{X is the number of layers} & 2, 2, 0.8 & $0.9\times 10^{6}$ & $0.8\times 10^6$ & $1.5\times 10^{-11}$ & $1.0\times 10^{-12}$ & $5\times 10^{-13}$ & 0.15 \\
    \hline
    \end{tabular}
    }
    \caption{Details of simulated parameters}
    \label{tab:table1}
\end{table}

\begin{acknowledgement}
The authors would like to thank King Abdullah University of Science and Technology for the funding support for this project.
\end{acknowledgement}

\begin{suppinfo}
The supporting data will be available on a reasonable request to the authors.
\end{suppinfo}

\subsection{Author Contribution}
A. H. Lone conceived the idea and fabricated the devices along with X. Zou. A. H. Lone did the characterization of the devices supported by V. S.  The micromagnetic simulations and neural network simulations were carried by A. L. K. M. worked on the post-processing of the simulation data. The paper was written by A. H. Lone with support from K. Mishra, G. Setti and H. Fariborzi. The project was supervised by H. Fariborzi and G. Setti. 

\bibliography{Ref-2}

\end{document}